\documentstyle[12pt]{article}
\newcommand{\eq}[1]{(\ref{#1})}
\newcommand{\formel}[1]{\begin{equation} #1 \end{equation}\noindent}
\newcommand{\formlab}[2]
{\begin{equation}\label{#1} #2 \end{equation}\noindent}

\begin{document}
\pagestyle{plain}
\parindent=10mm
\parskip=3mm

\baselineskip=30pt
\begin{center}
{\huge\bf Fast vectorized algorithm for the Monte Carlo Simulation
of the Random Field Ising Model}
\end{center}
\vskip0.5cm
\baselineskip=18pt
\begin{center}
{\LARGE H. Rieger\footnote{\normalsize
e--mail address: heiko@chromo.ucsc.edu or
heiko@hlrsun.hlrz.kfa--juelich.de}}
\end{center}
\begin{center}
{\Large
Physics Department\\
University of California\\
Santa Cruz, CA 95064, USA\\
and\\
HLRZ, c/o KFA J\"ulich\\
Postfach 1913\\
5170 J\"ulich, Germany\\}
\end{center}
\vskip2.5cm
\baselineskip=18pt
{\large\bf Abstract:} An algoritm for the simulation of the
3--dimensional random
field Ising model with a binary distribution of the random fields is
presented. It uses multi-spin coding and simulates 64 physically
different systems simultaneously. On one processor of a Cray YMP it
reaches a speed
of 184 Million spin updates per second. For smaller field strength
we present a version of the algorithm that can perform 242 Million
spin updates per second on the same machine.
\vskip2cm
\parindent=0mm
{\large\bf Key words:} Monte Carlo Simulation, Multi Spin Coding,
Random Field Ising Model.
\parindent=10mm
\vfill
\eject

\section{Introduction}
Still there are many open question in connection with the random field
Ising model --- RFIM --- (see ref.\ [1] for a recent review on this
subject). It has been shown rigorously that in more than two dimensions
the RFIM possesses a second order phase transition at finite
temperature for small enough field strength. Nevertheless there is
still much uncertainty concerning the exponents characterizing this
transition in three dimensions.
Results of computer--simulations and experimental data
seem to contradict each other if one tries to harmonize them with a
proposed scaling theory for it (see [1] for details). We have the
impression that more extensive simulation might help to clarify
this situation.

As a first step into this direction we present here an
effective algorithm that can perform Monte Carlo simulations of
the RFIM with a speed of 184 Milllion spin updates (MUPS) per second on one
processor of a Cray YMP.  It was developed out of the fast vectorized
algorithm for the simulation of the three dimensional Ising model,
which was originally invented by N.\ Ito [2], reaching a speed of 2190
MUPS on a Fujitsu VP 2600/10 and 800 MUPS on the NEC--supercomputer.
Later it was improved by H.\ O.\ Heuer [3] and implemented on a Cray
YMP, where it reached a speed of 305 MUPS on one processor of a Cray YMP.

The main idea we followed in the construction of
the code is to consider the two cases "spin parallel/antiparallel to
the external field" separately. Therefore (in the case of a binary
distribution for the fields) one cannot do worse than
double the time needed for the innermost loop in the RFIM in
comparison to the pure case considered in [3]. Taking into account the
fact that we use periodic boundary conditions instead of helical
or self consistent boundary conditions
(which gives a slowing down of approximately 10\%) our code has
surpassed this minimal requirement by about 30\%.
The speed of our algorithm has to be compared with the following data:
8 years ago the RFIM was simulated with a speed of 1 MUPS on a CDC 176
[4] and one year later the distributed array processor
(DAP) at Queen Mary College, London was able to update 14.6 Million
spins per second [5].

The exact definition of the model that the algorithm is able to simulate is
as follows. We consider a simple cubic lattice of linear dimension $L$
with $N=L*L*L$ Ising Spins $S_i=\pm1$. The Hamiltonian of the RFIM is
\formel{H=-J\sum_{\langle i\,j\rangle} S_i S_j-\sum_i h_i S_i\;,}
where the first sum extends over all nearest neighbor pairs
$\langle i\,j\rangle$ and the second sum over all sites. By rescaling
the temperature we confine ourselve to the case $J=1$. The external
fields $h_i$ are random variables obeying a binary probability
distribution
\formel{{\bf
P}(h_i)\,=\,p\cdot\delta(h_i-h)+(1-p)\cdot\delta(h_i+h)\;,}
where $p\in[0,1]$. Most of the literature deals with $p=1/2$.
Note that $p=1$ or $p=0$ yields the 3-d Ising model
in a homogeneous external field. For the version of the algorithm we
present in this
paper the field strength $h$ has to be smaller than 2, i.e.\
$h\in[0,2]$, for higher field values slight modifications have to be
incorporated. If one restricts oneself to field strength smaller than
one ($h\le1$), an even faster (242 Million spin--updates per second)
version of the algorithm can be used,
which is described in appendix A.

Since much of the code is a straightforward generalization of Heuer's
algorithm [3] we do not spend too much time in explaining the bulk of it.
Only the innermost loop has to be described in detail, which is done
in section 2. Furthermore we want to focus some attention on the
following point. The algorithm uses multispin coding and simulates 64
different systems at once. The essential speed--up in comparison to
older algorithms is achieved by using the same random number for
several systems. In the 3--d Ising model one has to simulate the 64
systems at different temperatures, otherwise one would run into
difficulties, since all 64 systems are identical, apart from the
initial condition. In our case, the RFIM, we deal with 64 physically
different systems, because of the different realizations of the
disorder, i.e.\ random field configurations. This is very convenient,
since at the end of the simulations we have to perform the average
over many disorder realizations anyway. Therefore we choose the same
temperature for each system and collect at the end the data for
magnetization, energy etc., until the desired number of realizations (most
conveniently a multiple of 64) is reached.

A few words to the whole program: First one initializes the random
number generator (which is a shift--register RNG \`a la Tausworth
[6]), then the initial spin--configurations of all 64 systems (each
bit in a computer word corresponds to one spin in one system). Now the
random field configurations are generated --- a bit in a computer
word is one if the random field at a particular site of one system
is positive, otherwise it is zero. Furthermore one has to initialize
the demon--arrays described below and also the nearest--neighbor
arrays. We use periodic boundary conditions since then finite size scaling
is expected to be more easy. To achieve vectorization one has to split
the whole system into two sublattices of size $N/2$. Therefore the
linear dimension $L$ of the cube has to be an even number.

After the initializations everything is set to perform
the update algorithm according to Metropolis, where a spin is
flipped with a probability
\formel{{\bf w}_{{\rm flip}} = {\rm min}\,\{\exp(-\beta\Delta
E),\,1\}\;,\quad \Delta E = E(-S_i)-E(S_i)\;.}
The subroutine {\bf sweep} described in section 2 updates
sequentially all spins in one sublattice. Hence for each MC--step
the routine {\bf sweep} is called twice: once for the update of the
first sublattice and once for the second. Therefore one needs also two
different nearest--neighbor arrays for the $x$--direction in the
different sublattices. For measurements one
needs an effective bit--counting routine, which can be found in [3].

\section{Description of the innermost loop}

The innermost loop is listed in figure 1. It is written in Fortran 77
and vectorizes on the Cray YMP. We tried to use a similar notation as
in the literature [2,3,7], to make comparisons easier.
Because we use a recursive algorithm
to generate the random numbers the compiler directive "ignore vector
dependency" {\tt IVDEP} occurs. Since the vector length of the
Cray computer systems is 64 this directive has no influence on the
correctness of the code (see [3] for details).

The subroutine {\bf sweep} updates the spins {\tt s(i)} (${\tt
i}=1,\ldots,{\tt N}$) of one sublattice, therefore {\tt N} equals
one half the total number of spins. The neighboring spins in the
second sublattice are stored in {\tt sneighb()}, and the random
fields are stored in {\tt h()}. The integers {\tt nxm(i)}, {\tt nxp(i)}
{\tt nym(i)}, {\tt nyp(i)}, {\tt nzm(i)} and {\tt nzp(i)} are the
indices from the six nearest neighbors of {\tt s(i)} (note that we use
periodic boundary conditions).
The array {\tt ir()} contains 256 random integers between 0 and {\tt irlst},
each composed of {\tt irbit} random bits. It is the essential part
of the shift--register random number generator.
The function of the demon--arrays {\tt ixp1()}, {\tt ixp2()}, {\tt
ixp3()}, {\tt ixm1()} and {\tt ixm2()} will be described later.

Let us assume that the arrays {\tt s()}, {\tt sneighb()} and {\tt
h()} are bit--arrays --- ignoring the fact that we deal with 64
bits at once and in parallel. The six bits ${\tt i1},\ldots,{\tt i6}$
(lines 13--18) contain the information about the orientation of spin
{\tt s(i)} with respect to its six neighbors. The bit is one if {\tt
s(i)} is antiparallel to the corresponding neighbor and it is zero
otherwise. The bit {\tt ih} (line 19) is one if the spin {\tt s(i)}
and the random field {\tt h(i)} are pointing in the opposite
direction.

In lines 20--30 the number of antiparallel pairs $\Sigma={\tt
i1}+\ldots+{\tt i6}$ is calculated.
The summation is done by adding {\tt i1}, {\tt i2} and {\tt i3} first
and storing the result in binary code into the two bits {\tt j1} and
{\tt j2} (${\tt i1}+{\tt i2}+{\tt i3}=2^1*{\tt j2}+2^0*{\tt
j1}$, see lines 20--22) and then adding {\tt i4}, {\tt i5} and
{\tt i6}, which yields {\tt j3} and {\tt j4} (see lines 23--25). In lines
26--30 the results of these two summation is added and since it is a
number between 0 and 6 one needs three bits $({\tt b3},{\tt b2},{\tt
b1})$ to store this information in binary code. Their meaning is given by
$\Sigma=2^2*{\tt b3}+2^1*{\tt b2}+2^0*{\tt b1}$.

The random number {\tt irt}, which is needed for the update of
all 64 spins {\tt s(i)} in the different systems, is generated
in lines 31--34. To save memory the random numbers lie on a wheel with
256 spokes, which is enforced by the periodic neighbor array {\tt
irndx()} (see [3] for details).

To make a flip--decision one has now to consider the possibilities ${\tt
ih}=0$ (spin and random field pointing into the same direction) and
${\tt ih}=1$ (spin and random field pointing in opposite directions)
separately. The case ${\tt ih}=0$ is considered first (lines 35--41).
A look at table 1 may help to explain the procedure. There we defined
$p_0=\exp[-(12+2h)\beta]$, $p_1=\exp[-(8+2h)\beta]$,
$p_2=\exp[-(4+2h)\beta]$ and $p_3=\exp[-2h\,\beta]$. One reads the
table in the following way: If for instance $\Sigma=0$, all six
nearest neighbors are parallel to spin {\tt s(i)} and the energy
difference to the state with {\tt s(i)} flipped would be $\Delta E =
12+2h$, which means that {\tt s(i)} should be flipped with a
probability $p_0=\exp[-(12+2h)\beta]$. Analogously for
$\Sigma=1,\ldots,6$.
\begin{center}
\begin{tabular}{l|@{$\quad$}r@{$\quad$}|c|c}
$\Sigma$ & $\Delta E$ & flip--prob. & $ixp$ \\ \hline
$0$ & $12+2h$ & $p_0$ & $4$ \\
$1$ & $8+2h$ & $p_1$ & $3$ \\
$2$ & $4+2h$ & $p_2$ & $2$ \\
$3$ & $2h$ & $p_3$ & $1$ \\ \hline
$4$ & $<0$ & $1$ & $0$ \\
$5$ & $<0$ & $1$ & $0$ \\
$6$ & $<0$ & $1$ & $0$ \\ \hline
\end{tabular}
\end{center}
\begin{center}
{\bf Table 1}
\end{center}
Here it is important that the field strength $h$ is not greater than 2,
otherwise also the energy differnce $\Delta E$ in the case $\Sigma=4$
would be positive ($\Delta E = -4+2h > 0$ for $h>2$), which would
result in a flip probability smaller than one. Nevertheless for $h>2$
the algoritm can easily be modified. In fact one has to change the
algorithm for each of the cases $h\in[2,4]$, $h\in[4,6]$ and
$h>6$ in a different way, but for physical reasons field strength
larger than two are not advisable to simulate.

Now we construct a flip--bit {\tt id0} by adding to the sum $\Sigma$
a demon--number $ixp\in\{0,1,\ldots,4\}$, which is composed by
3 demon--bits {\tt ix1},
{\tt ix2} and {\tt ix3} via $ixp=2^2*{\tt ix3}+2^1*{\tt ix2}+2^0*{\tt ix1}$.
If $(\Sigma+ixp)\ge4$, then the spin has to be flipped, which means ${\tt
id0}=1$, otherwise ${\tt id0}=0$. Certainly ${\tt id0}=1$ if
{\tt b3} is one (line 40), also if {\tt ix3} is one (line 41).
If {\tt b3} and {\tt ixp3} are both zero there is still the possibility
that the following addition yields an overflow bit:
\formlab{add}{
\begin{tabular}{r r r r r}
 & ( & {\tt b2} & {\tt b1} & ) \\
 & + ( & {\tt ix2} & {\tt ix1} & ) \\ \hline
(& {\tt id0} & $\bullet$ & $\bullet$ & ) \\
\end{tabular}
}
which is done in lines 38--39. First (line 38) one checks, whether the
addition
of {\tt b1} and {\tt ix1} is greater than 1, which yields an overflow
bit for the sum of the lowest bits if the two numbers $\Sigma$ and
$ixp$. Then (line 39) one takes this overflow bit and adds it to the sum of
{\tt b2} and {\tt ix2}. This gives an overflow bit for the sum of the
two smallest bits of each of the two  numbers $\Sigma$ and $ixp$.
If this overflow
bit is one it means that the sum is greater than 4, as desired.

{}From table 1 we also learn with which
probabilities the demon--number $ixp$ has to be set to the different
values: it has to be 4 with probability $p_0$, 3 with probability
$p_1-p_0$, 2 with probability $p_2-p_1$, 1 with probability $p_3-p_2$
and 0 with probability $1-p_3$. If one works with a computer that has
only a very limited memory capacity one can operate with
if--instructions in an obvious way --- here we use a time saving trick
(but also memory gobbling, see [3]): We define the arrays
{\tt ixp1()}, {\tt ixp2()} and
{\tt ixp3()} as shown in table 2, where ${\tt irlst}=2^{{\tt irbit}}-1$ and
$k_0={\tt irlst}* p_0$, $k_1={\tt irlst}*(p_1-p_0)$,
$k_2={\tt irlst}*(p_2-p_1)$, $k_3={\tt irlst}*(p_3-p_2)$. Note that
the number of bits {\tt irbit} determines the accuracy of the
probabilities: here they have only {\tt irbit} significant bits and
therefore {\tt irbit} should not be smaller than 17.
\begin{center}
\begin{tabular}{c|c c c|c}
{\tt irt} & {\tt ixp3(irt)} \vline &  {\tt ixp2(irt)} \vline &
{\tt ixp1(irt)} &
$\hat=\,ixp$ \\\hline
$0,\ldots,k_0-1$ & 1 & 0 & 0 & $4$\\
$k_0,\ldots,k_1-1$ & 0 & 1 & 1 & $3$\\
$k_1,\ldots,k_2-1$ & 0 & 1 & 0 & $2$\\
$k_2,\ldots,k_3-1$ & 0 & 0 & 1 & $1$\\
$k_3,\ldots,\,${\tt irlst} & 0 & 0 & 0 & $0$\\ \hline
\end{tabular}
\end{center}
\begin{center}
{\bf Table 2}
\end{center}
Remember that the shift--register random number generator in line 33
produces an integer {\tt irt} uniformly distributed between 0 and {\tt
irlst} and therefore $ixp=2^2*{\tt ixp3(irt)}+2^1*{\tt
ixp2(irt)}+2^0*{\tt ixp1(irt)}$ has the desired feature to be four
with probability $p_o$, three with probability $(p_1-p_0)$, etc.

Now that we have calculated the flip--bit that has to be used in the
case ${\tt ih}=0$ we come to the construction of the flip--bit {\tt
id1} for the possibility ${\tt ih}=1$. First we compute
$\Sigma'=\Sigma+1$ (see lines 42--44 --- the reason for this will soon
be clear) and overwrite the bits {\tt b3}, {\tt b2} and {\tt b1} with
this result in binary code. Now we might need table 3, where
$p'_0=\exp[-(12-2h)\beta]$, $p'_1=\exp[-(8-2h)\beta]$ and
$p'_2=\exp[-(4-2h)\beta]$.
\begin{center}
\begin{tabular}{l|@{$\quad$}r@{$\quad$}|c|c}
$\Sigma'$ & $\Delta E$ & flip--prob. & $ixm$ \\ \hline
$1$ & $12-2h$ & $p'_0$ & $3$ \\
$2$ & $8-2h$ & $p'_1$ & $2$ \\
$3$ & $4-2h$ & $p'_2$ & $1$ \\ \hline
$4$ & $<0$ & $1$ & $0$ \\
$5$ & $<0$ & $1$ & $0$ \\
$6$ & $<0$ & $1$ & $0$ \\
$7$ & $<0$ & $1$ & $0$ \\ \hline
\end{tabular}
\end{center}
\begin{center}
{\bf Table 3}
\end{center}
Once again it is important that the field strength $h$ is smaller than 2,
otherwise the probability $p_2'$ would be greater than one.
The table has to be read in complete analogy to table 1.
Again we add a demon number $ixm$ to
$\Sigma'$ and the flip--bit {\tt id1} is one if
$(\Sigma'+ixm)\ge4$, which is certainly the case if ${\tt b3}=1$ (line 49).
Again {\tt id1} is also one if the same addition as in \eq{add} yields an
overflow--bit (lines 47--48).

The arrays {\tt ixm1()} and {\tt ixm2()} are defined as shown in table
4, where $k'_0={\tt irlst}* p'_0$, $k'_1={\tt irlst}*(p'_1-p'_0)$ and
$k'_2={\tt irlst}*(p'_2-p'_1)$. Note that in contrast to the case
${\tt ih}=0$ here $ixm$ ranges only from $0$ to $3$ (which saves
one array of length {\tt irlst}) since we added already
a one to $\Sigma$.
\begin{center}
\begin{tabular}{c|c c|c}
{\tt irt} &  {\tt ixm2(irt)} \vline & {\tt ixm1(irt)} &
$\hat=\,ixm$ \\\hline
$0,\ldots,k'_0-1$ & 1 & 1 & $3$\\
$k'_0,\ldots,k'_1-1$ & 1 & 0 & $2$\\
$k'_1,\ldots,k'_2-1$ & 0 & 1 & $1$\\
$k'_2,\ldots,\,${\tt irlst} & 0 & 0 & $0$\\ \hline
\end{tabular}
\end{center}
\begin{center}
{\bf Table 4}
\end{center}
\baselineskip17pt
Now we construct the proper spin--flip bit {\tt id} (line 50) via
\formel{
{\tt id}\,=\,\left\lbrace
{\begin{array}{ccccc}
{\tt id0} & \quad & {\rm if} & \quad & {\tt ih}=0\;,\\
{\tt id1} & \quad & {\rm if} & \quad & {\tt ih}=1\;,
\end{array}}\right.}
which is slightly faster than {\tt id = xor(id0,and(ih,xor(id0,id1)))}
proposed in [8] for the same operation. Finally the
spin {\tt s(i)} is flipped if and only if {\tt id} is one (line 51).
Instead of completing a cycle on the above mentioned wheel with 256
spokes for the RNG after each sweep through one sublattice we use a
counter {\tt icounter} that starts the wheel at the correct position
each time the subroutine {\bf sweep} is called.

\section{Summary}
We presented a fast vectorized algoritm for the MC--simulation of the
three--dimensional random field Ising model. We would like to point
out that this
algorithm is also very fast on non--vector machines. In this case one
can also dispense with the tricks that have to used to achieve
vectorization (like the two sublattices). By making the necessary
modifications for 32--bit integers one can implement it also very
efficiently on special--purpose computers, parallel computers and
transputer systems. In fact this is the way we used it to perform
very extensive simulations of the RFIM --- the results of this
investigation will be reported elsewhere [9].

Let us only mention that
the equilibration time of the random field Ising systems depends
strongly on the strength of the external fields. For small fields
(where one can use the faster version of the algorithms peresented in
this paper) the equilibration in finite systems is nearly as fast as
in the pure Ising model --- but very strong cross--over effects are
expected and the separation of the critical behavior of the disordered
system from the pure system is difficult. Therefore we prefered
higher field strength (where one has to use the slightly
slower algorithm) --- but in this case the equilibration of the
systems takes much longer, even for small system sizes. This makes
the need of a fast algorithm for the MC--simulation of the RFIM
obvious --- at least as long no {\em efficient} cluster--algorithm is
available.

\baselineskip18pt
\section*{Acknowledgement}
I am indepted to A.\ P.\ Young for focusing my attention on
the RFIM and the problem this paper is concerned with. Furthermore I
would like to thank H.\ O.\ Heuer for sending me the complete code
of his algorithm described in [3] and for some useful correspondence.
I am also grateful to the San Diego Supercomputer Center and the
HLRZ at the KFA J\"ulich for the allocation of computer time.
This work was financially supported by the Deutsche
Forschungsgemeinschaft (DFG).

\section*{Appendix}

Here we present a version of the algorithm that ca be used in the
case of $h\le1$ and which is about 31\% faster. The reason of this
speed--up lies in the fact that one needs not to consider the two
cases ${\tt ih}=0$ and ${\tt ih}=1$ separately. In fact for the version
we want to present one has to replace lines 38--50 of the code in figure
1 by the following four lines:
{\tt
\begin{tabbing}
\hskip2cm\=123456\= \kill
{\bf 38'}\> \>    id = or ( ih, ix1 ) \\
{\bf 39'}\> \>    id = xor ( and(b1,ix2) , and( id, xor(b1,ix2) ) ) \\
{\bf 40'}\> \>    id = xor ( and(b2,ix3) , and( id, xor(b2,ix3) ) ) \\
{\bf 41'}\> \>    id = or ( id, b3 )
\end{tabbing}
}
The arrays {\tt ixm1()} and {\tt ixm2()} are not needed anymore.
Also the variable {\tt id0} and {\tt id1} are superfluous. To explain
the above modifications we take a look at table 5, where we have
defined
\formel{\tilde\Sigma=2*\Sigma+{\tt ih}+1}
and ${\tilde p}_{\tilde\Sigma}=
\exp[-\beta\Delta E(\tilde\Sigma)]$.
\begin{center}
\begin{tabular}{c|@{$\quad$}r@{$\quad$}|c|c}
$\tilde\Sigma$ & $\Delta E$ & flip--prob. & $ix$ \\ \hline
$1$ & $12+2h$ & ${\tilde p}_1$ & $7$ \\
$2$ & $12-2h$ & ${\tilde p}_2$ & $6$ \\
$3$ & $8+2h$ & ${\tilde p}_3$ & $5$ \\
$4$ & $8-2h$ & ${\tilde p}_4$ & $4$ \\
$5$ & $4+2h$ & ${\tilde p}_5$ & $3$ \\
$6$ & $4-2h$ & ${\tilde p}_6$ & $2$ \\
$7$ & $2h$ & ${\tilde p}_7$ & $1$ \\ \hline
$8$ & $<0$ & $1$ & $0$ \\
\vdots &\vdots &\vdots &\vdots \\
$14$ & $<0$ & $1$ & $0$ \\ \hline
\end{tabular}\\
\end{center}
\begin{center}
{\bf Table 5}\\
\end{center}
We observe that only as long as $h\le1$ the flip--probabilities
${\tilde p}_{\tilde\Sigma}$  are monotonically increasing with
$\tilde\Sigma$, which is necessary for the algorithm to work.
To construct the flip--bit {\tt id} we add to
$\tilde\Sigma$ a demon--number $ix\in\{0,1,\ldots,7\}$,
which is composed of 3 demon bits {\tt ix1}, {\tt ix2} and {\tt ix3}
via $ix=2^2*{\tt ix3}+2^1*{\tt ix2}+2^0*{\tt ix1}$. If
$(\tilde\Sigma+ix)\ge8$ the spin has to be flipped, meaning ${\tt
id}=1$. Obviously ${\tt id}=1$ if ${\tt b3}=1$ (line 41'). The only
other possibility for {\tt id} to become one is an overflow--bit on
the following addition:
\begin{equation}
\begin{tabular}{r r r r r r r}
 &   & ( & {\tt b2} & {\tt b1} & {\tt ih} & ) \\
 & + & ( & {\tt ix3} & {\tt ix2} & {\tt ix1} & ) \\
 & + & & & & 1 & \\ \hline
 & ( & {\tt id} & $\bullet$ & $\bullet$ & $\bullet$ & ) \\
\end{tabular}
\end{equation}
which is done in lines 38'--40'. The first term in this sum corresponds
to the lower three bits of $\tilde\Sigma$ in binary representation ---
the bits {\tt b2}, {\tt b1} are shifted to the left because of the
multiplication of $\Sigma$ with two and the lowest bit is then
occupied by {\tt ih} since it has
to be added. Furthermore in the definition of $\tilde\Sigma$ we added
a one, which is the origin of the last term in the sum. The arrays
{\tt ixp1()}, {\tt ixp2()} and {\tt ixp3()} have to be defined as depicted in
table 6, where ${\tilde k}_\nu={\tt irlst}*({\tilde p}_\nu-{\tilde
p}_{\nu-1})$, (${\tilde p}_0=0$).
\begin{center}
\begin{tabular}{c|c c c|c}
{\tt irt} & {\tt ixp3(irt)} \vline &  {\tt ixp2(irt)} \vline &
{\tt ixp1(irt)} &
$\hat=\,ix$ \\\hline
$0,\ldots,{\tilde k}_1-1$ & 1 & 1 & 1 & $7$\\
${\tilde k}_1,\ldots,{\tilde k}_2-1$ & 1 & 1 & 0 & $6$\\
$\vdots$ & $\vdots$ & $\vdots$ & $\vdots$ & $\vdots$\\
${\tilde k}_7,\ldots,\,${\tt irlst} & 0 & 0 & 0 & $0$\\ \hline
\end{tabular}
\end{center}
\begin{center}
{\bf Table 6}
\end{center}
This modification of the demon arrays has of course to be done in the
initialization procedures.
\vfill
\eject

\section*{References}

\baselineskip=15pt

\newcommand{\lll}[4]{ #1 {\bf{#2}} (19#3) #4.}
\newcommand{\z}{{\em Z.\ Phys.\ }}
\newcommand{\jp}{{\em J.\ Phys.\ }}
\newcommand{\jphys}{{\em J.\ Physique}}
\newcommand{\pr}{{\em Phys.\ Rev.\ }}
\newcommand{\prl}{{\em Phys.\ Rev.\ Let.\ }}
\newcommand{\js}{{\em J.\ Stat.\ Phys.\ }}

\newcounter{lit}
\begin{list}
{[\arabic{lit}]}{\usecounter{lit}\setlength{\rightmargin}{\leftmargin}}

\item D.\ P.\ Belanger and A.\ P.\ Young, \lll {{\em Journal
of Magnetism and Magnetic Materials}} {100} {91} {272}

\item N.\ Ito and Y.\ Kanada, \lll {{\em Supercomputer}} {25} {88} {31}

\item H.\ O.\ Heuer, \lll {{\em Comput.\ Phys.\ Commun.\ }} {59} {90}
{387}

\item D.\ Stauffer, C.\ Hartzstein, K.\ Binder and A.\ Aharony, \lll {\z}
{B55} {84} {325}

\item A.\ P.\ Young and M.\ Nauenberg, \lll {\prl} {54} {85} {2429}

\item R.\ Tausworth, \lll {{\em Math.\ Comput.\ }} {19} {65} {201}

\item E.\ Bhanot, D.\ Duke and R.\ Salvador, \lll {\js} {44} {86} {85}

\item F.\ Bagnoli, \lll {{\em Int.\ J.\ Mod.\ Phys.\ }} {3} {91} {307}

\item H.\ Rieger and A.\ P.\ Young, in preparation.

\end{list}

\vfill
\eject
\begin{center}
\underline{\large{\bf Figure 1}}
\end{center}
\hskip0.5cm
{\tt\baselineskip=10pt
\begin{tabbing}
\hskip2cm\=123456\= \kill
{\bf 1}\> \>     {\bf subroutine sweep}
(s,sneighb,h,nxm,nxp,nym,nyp,nzm,nzp,N)\\
\> \> \\
{\bf 2}\> \>     integer s(N),sneighb(N),h(N)\\
{\bf 3}\> \>     integer nxm(N),nxp(N),nym(N),nyp(N),nzm(N),nzp(N)\\
{\bf 4}\> \>     integer spin,b1,b2,b3\\
{\bf 5}\> \>     parameter (irbit=18)\\
{\bf 6}\> \>     parameter (irlst = 2**irbit-1)\\
{\bf 7}\> \>     common /rnddim/ ir(0:255),irndx(2,0:255),icounter\\
{\bf 8}\> \>     common /rngdl/  ixp1(0:irlst),ixp2(0:irlst),ixp3(0:irlst),\\
{\bf 9}\> \>    \hskip-5pt 1\hskip3.24cm ixm1(0:irlst),ixm2(0:irlst)\\
\> \> \\
{\bf 10}\> CDIR\$ IVDEP \> \\
\> \> \\
{\bf 11}\> \>      do 100 i = 1,N\\
\> \> \\
{\bf 12}\> \>    spin = s(i)\\
{\bf 13}\> \>    i1 = xor( spin, sneighb(nxm(i)) )\\
{\bf 14}\> \>    i2 = xor( spin, sneighb(nxp(i)) )\\
{\bf 15}\> \>    i3 = xor( spin, sneighb(nym(i)) )\\
{\bf 16}\> \>    i4 = xor( spin, sneighb(nyp(i)) )\\
{\bf 17}\> \>    i5 = xor( spin, sneighb(nzm(i)) )\\
{\bf 18}\> \>    i6 = xor( spin, sneighb(nzp(i)) )\\
{\bf 19}\> \>    ih = xor( spin, h(i) )\\
\> \> \\
{\bf 20}\> \>    j2 = xor( i1, i2 )\\
{\bf 21}\> \>    j1 = xor( j2, i3 )\\
{\bf 22}\> \>    j2 = xor( and( i1, i2), and( j2, i3 ) )\\
{\bf 23}\> \>    j4 = xor( i4, i5 )\\
{\bf 24}\> \>    j3 = xor( j4, i6 )\\
{\bf 25}\> \>    j4 = xor( and( i4, i5), and( j4, i6 ) )\\
\> \> \\
{\bf 26}\> \>    b1 = and( j1, j3 )\\
{\bf 27}\> \>    b3 = xor( j2, j4 )\\
{\bf 28}\> \>    b2 = xor( b3, b1 )\\
{\bf 29}\> \>    b3 = xor( and( j2, j4 ), and( b3, b1) )\\
{\bf 30}\> \>    b1 = xor( j1, j3 )\\
\> \> \\
{\bf 31}\> \>    index = i + icounter\\
{\bf 32}\> \>    j     = and(  index, 255 )\\
{\bf 33}\> \>    irt   = xor( ir(irndx(1,j)), ir(irndx(2,j)) )\\
{\bf 34}\> \>    ir(j) = irt\\
\> \> \\
{\bf 35}\> \>    ix1 = ixp1(irt)\\
{\bf 36}\> \>    ix2 = ixp2(irt)\\
{\bf 37}\> \>    ix3 = ixp3(irt)\\
{\bf 38}\> \>    id0 = and( b1, ix1 )\\
{\bf 39}\> \>    id0 = xor( and(b2,ix2), and(xor(b2,ix2),id0) )\\
{\bf 40}\> \>    id0 = or( id0,  b3 )\\
{\bf 41}\> \>    id0 = or( id0, ix3 )\\
\> \> \\
{\bf 42}\> \>    b3 = or( b3, and( b2, b1) )\\
{\bf 43}\> \>    b2 = xor( b2,  b1 )\\
{\bf 44}\> \>    b1 = not ( b1 )\\
\> \> \\
{\bf 45}\> \>    ix1 = ixm1(irt)\\
{\bf 46}\> \>    ix2 = ixm2(irt)\\
{\bf 47}\> \>    id1 = and( b1, ix1 )\\
{\bf 48}\> \>    id1 = xor( and(b2,ix2), and(xor(b2,ix2),id1) )\\
{\bf 49}\> \>    id1 = or( id1, b3 )\\
\> \> \\
{\bf 50}\> \>    id = or ( and(id0,not(ih)) , and(id1,ih) )\\
\> \> \\
{\bf 51}\> \>    s(i) = xor( s(i), id )\\
\> \> \\
{\bf 52}\> \hskip10pt 100 \> continue\\
\> \> \\
{\bf 53}\> \>    icounter = icounter + N\\
\> \> \\
{\bf 54}\> \>    return\\
{\bf 55}\> \>    end
\end{tabbing}
}

\end{document}